# Red Blood Cell Sphericity Index Obtained by Defocusing Microscopy and Retinopathy Severity in Sickle Cell Disease

Camilo Brandão-de-Resende [1,2], Ubirajara Agero[3], Oscar N. Mesquita [3],

Lívia S. Gomes [3], Daniel V. Vasconcelos-Santos [1,2,4]

[1] Hospital São Geraldo / Hospital das Clínicas da Universidade Federal de Minas Gerais, Belo Horizonte, Brazil

[2] Programa de Pós-Graduação em Ciências da Saúde – Infectologia e Medicina Tropical, Universidade Federal de Minas Gerais, Belo Horizonte, Brazil.

[3] Departamento de Física, Universidade Federal de Minas Gerais, Belo Horizonte, Brazil.

[4] Departamento de Oftalmologia e Otorrinolaringologia, Universidade Federal de Minas Gerais, Belo Horizonte, Brazil.

Based on PhD thesis:

https://repositorio.ufmg.br/bitstream/1843/38353/1/TESE%20CAMILO%20-%20FINAL.pdf

Code:

https://github.com/camiloresende/defocusing_microscopy_R

# ABSTRACT

Proliferative sickle cell retinopathy (PSCR) is the most important ocular manifestation of sickle cell disease (SCD), but understanding of its pathophysiology remains incomplete. Red blood cell (RBC) deformability has been identified as a critical factor in SCD and is intrinsically related with cell sphericity index (SI). The aim of this study was to develop an algorithm that automatically analyzes defocusing microscopy (DM) images, providing biomechanical parameters of individual RBCs, including cellular SI. Furthermore, association of RBC SI with severity of PSCR in individuals with HbSC-SCD was also investigated. We determined the SI of 255 RBCs from 17 individuals using DM. Nine individuals (53%) were healthy and eight (47%) had HbSC-SCD. Mean SI was significantly lower in SCD than in control group (0.69 versus 0.73), and also in severe than in mild PSCR (0.67 versus 0.70). Lower SI was also associated with more severe retinopathy (ROC AUC = 0.85) and the proportion of participants with SI ≤ 0.69 was significantly greater in severe than in mild PSCR (100% versus 25%). In conclusion, lower RBC SI determined by DM was associated not only with the presence of SCD, but also with more severe PSCR, possibly being a surrogate biomarker of disease severity.

**Keywords**: Anemia, Sickle Cell; Light Microscopy; Hemoglobin SC Disease; Retina.

**Introduction**

Sickle cell disease (SCD) is a life-threatening hematological disorder that affects millions of people and is a global health problem, especially among afrodescendants [1,2]. It is caused by abnormal beta-globin alleles carrying inherited sickle mutation (Glu6Val) on the hemoglobin subunit beta gene, resulting in the hemoglobin S gene (HbS) [1,2]. The disease can present in the homozygous form HbSS, with inheritance of HbS from both parents, or in heterozygous forms, such as hemoglobin C (HbC) with HbS (HbSC) [2]. SCD affects multiple systems and virtually any organ may be involved [1,3]. Ocular manifestations of the disease, although common, are probably underdiagnosed because of chronic course and absence of symptoms until more advanced stages [3]. Proliferative sickle cell retinopathy (PSCR) is the most important ophthalmic manifestation of SCD, since peripheral retinal neovascularization and its sequelae can be vision threatening, especially when vitreous hemorrhage or retinal detachment ensues [3].

SCD presents a broad spectrum of clinical manifestations with variable severity, even among individuals with the same genotype (e.g. HbSS or HbSC), and its pathophysiology is complex and not completely understood [1-3]. For example, knowledge of factors associated with occurrence and severity of PSCR is scarce [4]. Risk factors for PSCR described in the literature include HbSC subtype, older age, male gender, lower levels of fetal hemoglobin (HbF), and higher hemoglobin concentration [3-7]. However, none of these factors can be consistently used to stratify individuals with SCD according to their risk or severity of PSCR.

Deformability reflects cell ability to change its shape under applied stress and is essential for RBCs to pass several times a day through narrow capillaries to release oxygen to the tissues [8,9]. RBC deformability has been identified as a critical factor for vascular occlusion in SCD and is correlated with severity of some disease manifestations, such as pain crises and frequency of microvascular occlusions [10-12]. A decrease in deformability was observed in RBCs in diabetic retinopathy, and may also be associated with other retinal vascular disorders [8,13]. Cell deformability is mainly determined by three factors: (1) volume to surface area ratio (i.e., cell sphericity), (2) internal viscosity, which is determined by the mean cell hemoglobin concentration (MCHC), and (3) rheological properties of the cell membrane [9]. The sphericity index is a dimensionless number, defined as a normalized volume-to-surface area ratio ($4.836 \times \text{Volume}^{2/3} / \text{Surface Area}$), with a value of one for perfect spheres and of zero for flat disks [14].

Defocusing microscopy (DM) is a quantitative phase imaging (QPI) technique [15-18] that uses defocus to observe phase objects like unlabeled RBCs [19-22]. The technique is simple and does not require use of any extra optical element, so that it can be easily adopted by non-specialists. Morphological, chemical and mechanical parameters of individual RBC can be measured by DM, such as cell sphericity index and membrane mean fluctuation amplitude [21], which are intrinsically related with RBC deformability [9]. Recently, DM intensity images were used to study morphological parameters of RBCs in retinal vascular disorders [13]. Furthermore, DM fluctuation intensity images have been applied to retrieve cells membrane fluctuations and mechanical parameters [23-25].

We here share the code of a novel algorithm to analyze DM intensity images using the software R v. 3.6.3 [26] (**Supplementary code**). The algorithm simultaneously analyzes DM images from multiple RBCs of an individual and determines several biomechanical RBC parameters, such as RBC radius, surface area, volume, sphericity index and membrane mean fluctuation amplitude at the center of the cell. We focused our study in analyzing RBC sphericity index and in evaluating its association with the presence of SCD and severity of PSCR in subjects with HbSC.

Previous algorithms involved image pre-processing and analysis using multiple software packages and took more than one hour per cell to compile results, especially to determine cell asymmetry [21]. The developed algorithm takes about seven seconds to simultaneously analyze multiple DM images of a cell, using a notebook computer (Intel® Core™ i7-7500U 2.70-2.9GHz, 8GB RAM, with Windows 10).

**Materials and Methods**

**Study design and participants**

This is an observational, case-control study to analyze associations of RBC sphericity index with presence of SCD and severity of PSCR. The study was approved by institutional review board of the Universidade Federal de Minas Gerais (CAAE 97392218.2.0000.5149) and was performed in accordance with the principles of the declaration of Helsinki, with written informed consent being obtained from all participants.

We analyzed 255 RBCs from 17 consecutive participants (15 RBCs per individual) to determine cell sphericity index using DM. Only individuals with age ≥ 18 years, without diabetes or systemic arterial hypertension, and with no history of blood transfusion in the previous six months were included. Nine participants (53%) were healthy (with HbAA) and eight (47%) had SCD, subtype HbSC. We only included individuals with HbSC (confirmed by hemoglobin electrophoresis) in the SCD group because this is the disease subtype with greatest prevalence of PSCR [3]. All participants underwent complete ophthalmic examination, including assessment of best-corrected VA (visual acuity reported in logMAR scale) with an ETDRS (Early Treatment of Diabetic Retinopathy Study) chart, applanation tonometry, slit-lamp examination (biomicroscopy), and indirect ophthalmoscopy. Demographic (age and gender) and laboratory data [ABO and Rh blood type, hemoglobin concentration (Hb) and mean corpuscular hemoglobin concentration (MCHC)] were also assessed for all participants. For those with SCD, we also evaluated reticulocyte count, current use of hydroxycarbamide (hydroxyurea) and history of disease complications, including pain crisis, acute thoracic syndrome, osteomyelitis, biliary complications, priapism and stroke. Only laboratory results obtained less than six months before blood sample collection and analysis were included.

Firstly, we examined demographic, clinical, and laboratory characteristics in participants without (control) and with SCD, and also with less severe (mild PSCR) and more severe PSCR (severe PSCR). Afterwards, we compared RBC sphericity index determined by DM (the average value for the 15 RBCs analyzed for each participant) between control and SCD and between mild PSCR and severe PSCR groups.

The most widely used severity scale for grading PSCR was proposed by Goldberg, and is composed by five stages [27]. Stage I consists of peripheral arteriolar occlusions, and stage II of peripheral arterio-venous anastomoses. Stage III is characterized by the classic 'sea-fan' neovascularization in the retinal periphery at the border of perfused and nonperfused retina. Stage IV represents vitreous hemorrhage that occurs when a frond of neovascularization bleeds. Finally, in stage V there is either tractional or tractional–rhegmatogenous retinal detachment. PSCR was considered as severe if staged ≥ III (presence of 'sea-fan' neovascularization) and otherwise as mild, accordingly to Goldberg's classification. All participants with severe PSCR underwent scatter laser treatment around the 'sea-fan' neovascularizations.

**Experimental Setup**

All blood samples were collected in 2mL EDTA tubes. Between two and three hours after collection, we diluted 0.5 µL of blood in 2 mL of phosphate-buffered saline (PBS) solution with bovine serum albumin (BSA) 1% (10mg/mL). All samples were analyzed within one hour after dilution. Only RBCs with normal appearance were included; those with signs of sickling, lysis or with intracellular content were excluded.

Experiments were performed using an inverted microscope (Nikon Eclipse TI-E, Nikon, Melville, New York) operating in bright-field mode and setup for Köhler illumination. Illumination was with a halogen lamp (100 W) and the diffracted rays were collected by an oil immersion objective (Nikon Plan APO DIC H, 100X, NA 1.49; Nikon). A long-pass filter was used to allow only transmission of wavelengths above 610 nm, since RBCs absorb light in the blue range of visible spectrum [21]. Videos were captured with a CMOS camera (Silicon Video CMOS 642M, 8 bits, Epix Inc., Buffalo Grove, Illinois) at 300 frames per second, with exposure time of 3.33 µs, and image gain to keep background gray level intensity between 120 and 130 (**Figure 1**). Sample temperature was controlled at 37ºC (ChamlideIC-CU:109, Live Cell Instrument, Nowan-gu, Korea). The defocus position $z_f = 0$ was set at the cell minimum contrast plane, and positive defocus positions ($z_f > 0$) were defined as above the cell minimum contrast plane and negative defocus positions ($z_f < 0$) as below it (**Figure 2**). We obtained images of background (300 frames) and of each RBC at $z_f = 0$ µm (300 frames), $z_f = +2$ µm (300 frames), and $z_f = +4$ µm (1000 frames). Images taken at $z_f = 0$ µm and $z_f = +2$ µm were used to determine RBC morphological parameters, while the images taken at $z_f = +4$ µm were used to determine the membrane mean fluctuation amplitude at the center of the RBC. Membrane fluctuation data were not the focus of this study, so these results are not presented here. All images (256x256 pixels) were saved in TIFF format. In average, 50 minutes were sufficient to acquire images of 15 cells from each participant, which were analyzed using the developed algorithm in less than 2 minutes.

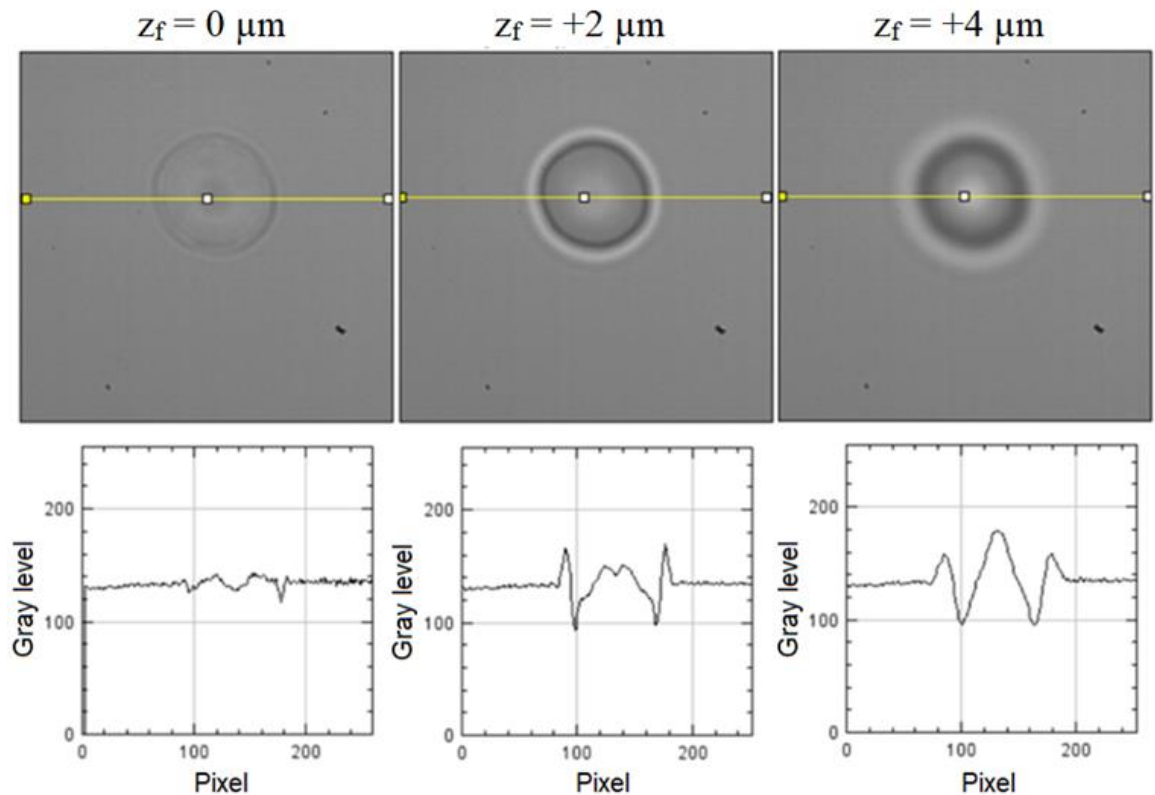

**Figure 1** – (Top) Intensity images of a RBC taken at defocus distances $z_f = 0$ µm, $z_f = +2$ µm, and $z_f = + 4$ µm. (Bottom) Gray level profile along the horizontal line represented in top images.

**Data Processing**

We developed an algorithm to automatically analyze DM images using the software R v. 3.6.3 [26] (**Supplementary code**). Image pre-processing and multiple RBC analysis were all performed by the same algorithm. As inputs, we enter the path to a folder containing all the images acquired for a single individual (including the background and three images for each RBC taken at defocus positions 0 µm, +2 µm, and +4 µm, respectively) and the value of the MCHC obtained by a complete blood count exam. The mathematical derivation of DM equations used in the algorithm can be found in **Supplementary Equations**.

The algorithm automatically calculates the following characteristics for each RBC (**Figure 2**): radius, surface area (AREA), volume, sphericity index, maximum slope at cell concavity (SMAX), radial distance between the cell center and the point of SMAX (DSMAX), maximum cell concavity (CMAX), central cell thickness (T0), maximum cell thickness (TMAX), radial distance between the cell center and the point of TMAX (DTMAX), and membrane mean fluctuation amplitude at the center of the cell $\langle u \rangle$. The radius of cell is defined at the minimum derivative (more negative) of angular average of cell thickness with respect to the distance from the center (**Figure 3**) [21].

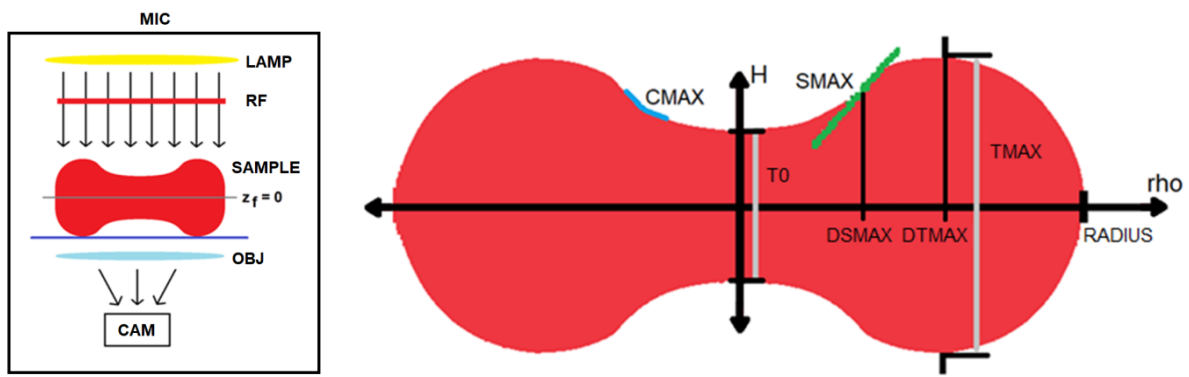

**Figure 2** – (left) DM setup: inverted optical microscope operating in bright-field mode (MIC), halogen lamp (LAMP), long-pass filter (RF), objective lens (OBJ), and camera (CAM). The defocus position $z_f = 0$ was set at the cell minimum contrast plane. (right) Schematic cross section of a RBC showing the defined reference points.

The algorithm returns as outputs the RBC characteristics and summary images of each analyzed cell (**Figure 3**). Also, it gives the average results of all analyzed cells from an individual (**Figure 4**). In order to accelerate the algorithm, we assumed cell symmetry along the x-axis. This assumption is valid for the purposes of this study because we are not interested in investigating the effects of cell adhesion to the substrate, which may result in small alterations of RBC characteristics [21].

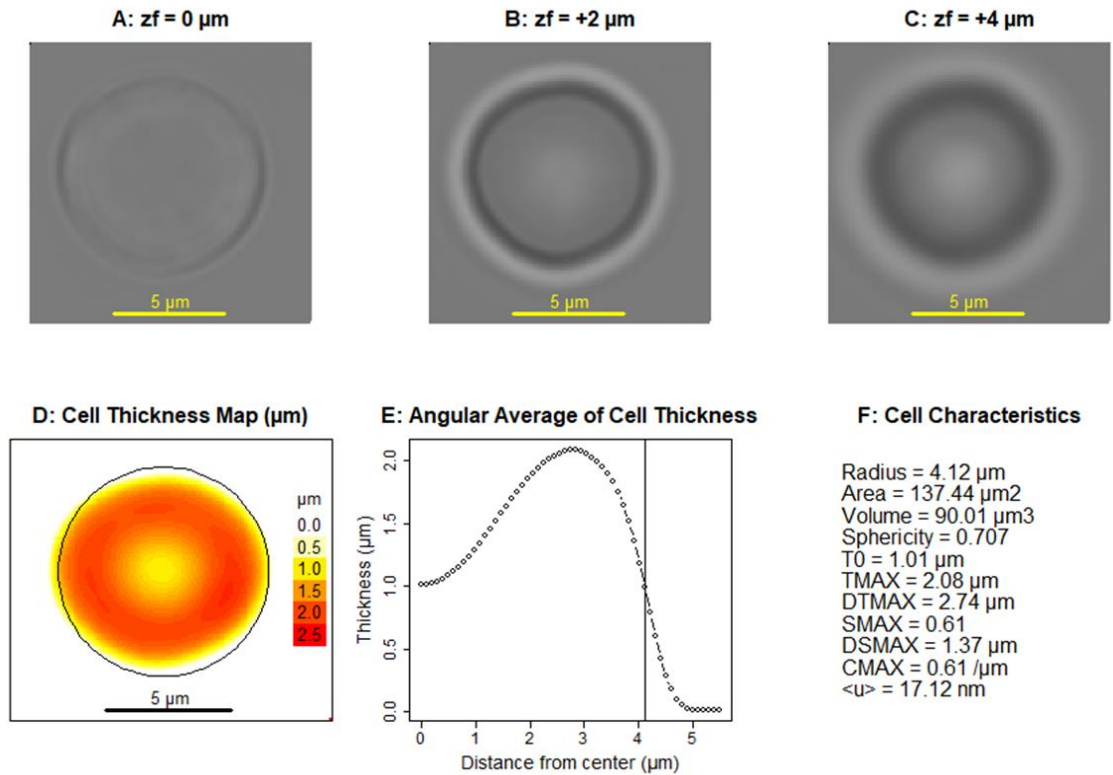

**Figure 3** – Summary of outputs returned by the algorithm. Mean intensity image of a cell at defocus distances $z_f = 0$ µm (A), $z_f = +2$ µm (B), and $z_f = +4$ µm (C). (D) Cell thickness map. (E) Angular average of the cell thickness. (F) Cell characteristics.

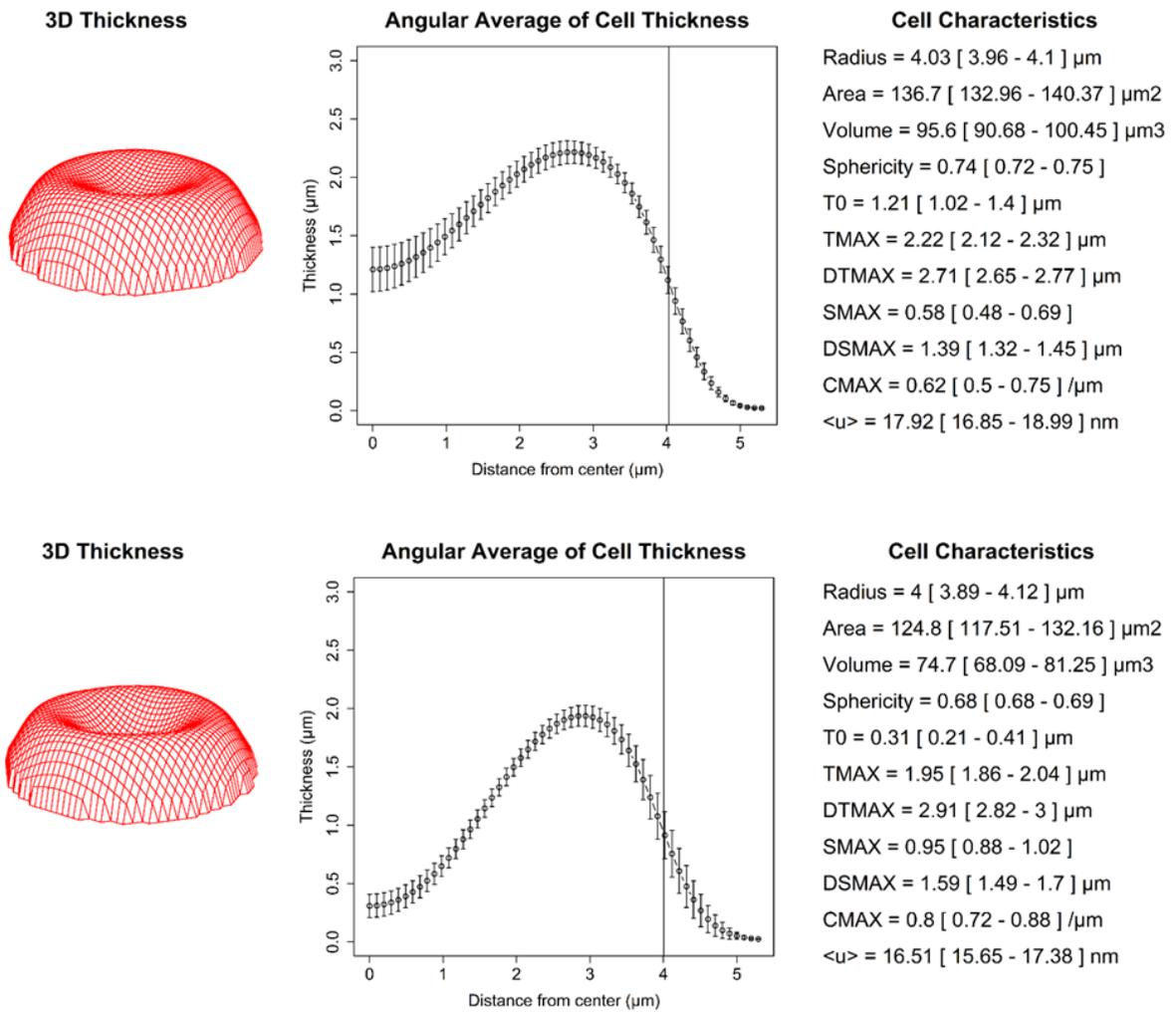

**Figure 4** – Average results for 15 cells analyzed from a healthy participant (top) and 15 cells analyzed from a participant with SCD (bottom): (left) 3D thickness profile, (center) cell thickness angular average, with vertical bars representing CI95% and (right) cell characteristics with average [CI 95%].

**Statistics**

As previously stated, sphericity index is directly related with RBC deformability [9]. In a pilot experiment analyzing 30 RBCs from a blood sample of a healthy participant, we estimated the mean sphericity index as 0.728 and its standard deviation as 0.035. In order to compare the means of two groups with the same size considering 90% test power and 5% significance level, each group should have at least five individuals [28]. We also estimated the number of cells to be analyzed for each individual in order to have the range of the 95%CI of estimated mean sphericity index lower than 5% of its value. To achieve this, we would need to analyze at least 15 RBCs per participant [28].

Statistical analyzes were performed using the software R v. 3.6.3 [26] with a significance level of 5% and using two-tailed tests. Data were described as mean [range] for continuous variables and as absolute number (percentage) for binary variables. Mean sphericity index was compared between groups using unequal variance t-tests after checking for normality in control and SCD groups using the Shapiro-Wilk test [29], which is the more appropriate normality test for small samples [30]. The area under ROC curve with 95%CI (bootstrap, 1,000 replicates) to predict severe PSCR from mean RBC sphericity index was also determined [31].

We also found a threshold of mean RBC sphericity index to predict severe PSCR, minimizing the sum of the Gini impurity within subgroups [32,33]. Gini impurity is a measure of statistical dispersion that is zero when all cases in a group have the same measured outcome and that increases up to a maximum value of one as the group becomes more heterogeneous. For example, in this study, subgroups in which none or all patients had severe PSCR would have a Gini impurity value of zero. Fisher exact test with a mid–P value adjustment was used to compare proportions of patients in subgroups and to estimate odds ratio (OR) confidence intervals [34].

**Results and Discussion**

We studied if RBC sphericity index was associated with SCD (HbSC), as well as with PSCR severity, analyzing 255 RBCs from 17 participants (15 RBCs per individual). Eight healthy participants composed the control group, while nine participants (HbSC) composed the SCD group. In the SCD group (N = 9), four individuals had mild PSCR, staged as Goldberg I, and five individuals had severe PSCR, staged as Goldberg III.

Only one eye of a participant with severe PSCR had visual acuity worse than 0.18, considered as normal vision (VA = 0.5 attributed to a macular epiretinal membrane). This result emphasizes the importance of periodic ophthalmic examinations in patients with SCD, since VA may be not affected even when advanced PSCR is present.

In **Table 1**, we summarize demographic, clinical, and laboratory data of the study population. We observed that gender, age, blood type and MCHC distributions were similar among all the subgroups analyzed. As expected, Hb levels were lower in individuals with SCD. Among participants with SCD, the most common previous complications of the disease were pain crises (100%), acute thoracic syndrome (78%), and osteomyelitis (11%). Pain crises are almost universal in SCD, and acute thoracic syndrome and osteomyelitis, in contrary to sickle cell retinopathy, are more common in HbSS than in HbSC, for reasons that are not completely understood [35].

Three participants with SCD (33%) were using hydroxycarbamide (hydroxyurea). This is a ribonucleotide reductase inhibitor with beneficial effects in SCD mainly due to an increase in fetal hemoglobin levels (HbF), inhibiting HbS polimerization [2]. Besides the effects on HbF, hydroxycarbamide also affects the transport of cations across the cell membrane, and RBC adhesion and morphology [36]. The drug increases the volume and sphericity index of RBCs [36,37]. In a recent study, no association was observed between hydroxycarbamide use and sickle cell retinopathy [4].

**Table 1** – Summary of demographic, clinical, and laboratory data in study population.

|  | Control (N=8) | SCD (N = 9) | | |
| --- | --- | --- | --- | --- |
|  |  | SCD Total (N = 9) | Mild PSCR (N=4) | Severe PSCR (N = 5) |
| **Female gender** | 6 (75) | 5 (56) | 2 (50) | 3 (60) |
| **Age (years)** | 28.5 [24 – 35] | 26.7 [18 – 43] | 26 [19 – 43] | 27 [18 – 43] |
| **VA in the worse eye (logMAR)** | 0.0 [0.0 – 0.0] | 0.0 [0.0 – 0.5] | 0.0 [0.0 – 0.0] | 0.1 [0.0 – 0.3] |
| **Blood type ABO / Rh** | A +: 5 (63) <br> B +: 2 (25) <br> B –: 0 (0) <br> O +: 1 (12) <br> O –: 0 (0) | A +: 4 (45) <br> B +: 1 (11) <br> B –: 1 (11) <br> O +: 1 (11) <br> O –: 2 (22) | A +: 1 (25) <br> B +: 0 (0) <br> B –: 1 (25) <br> O +: 0 (0) <br> O –: 2 (50) | A +: 3 (60) <br> B +: 1 (20) <br> B –: 0 (0) <br> O +: 1 (20) <br> O –: 0 (0) |
| **Hb (g/dL)** | 13.6 [12.0 – 15.0] | 11.2 [10.0 – 12.5] | 11.5 [10.5 – 12.5] | 11.0 [10.0 – 12.1] |
| **MCHC (g/dL)** | 33.6 [32.4 – 34.0] | 34.1 [33.0 – 36.7] | 33.9 [33.0 – 34.5] | 34.4 [33.1 – 36.7] |
| **Reticulocyte count (%)** | – | 3.5 [3.0 – 4.0] | 3.5 [3.0 – 4.0] | 3.5 [3.0 – 4.0] |
| **Hydroxycarbamide** | – | 3 (33) | 1 (25) | 2 (40) |
| **Pain crisis** | – | 9 (100) | 4 (100) | 5 (100) |
| **Acute thoracic syndrome** | – | 7 (78) | 3 (75) | 4 (80) |
| **Osteomielitis** | – | 1 (11) | 1 (25) | 0 (0) |
| **Biliary complications** | – | 0 (0) | 0 (0) | 0 (0) |
| **Priaprism** | – | 0 (0) | 0 (0) | 0 (0) |
| **Stroke** | – | 0 (0) | 0 (0) | 0 (0) |

**SCD:** sickle cell disease.
**Mild PSCR:** proliferative sickle cell retinopathy staged as Goldberg < III.
**Severe PSCR:** proliferative sickle cell retinopathy staged as Goldberg ≥ III.

In order to validate the algorithm, we compared the parameters (mean ± standard deviation) returned for the control group with those available in the literature. The measured RBC parameters were: diameter (7.9 ± 0.21) µm, maximum thickness (2.0 ± 0.2) µm, surface area (126.4 ± 8.3) µm$^2$, and volume (83.1 ± 9.7) µm$^3$. All these values were compatible with those reported in literature for healthy individuals [9,19]. In **Table 2**, we present the values of RBC radius, maximum thickness, surface area, volume, and sphericity index for Control and SCD groups. Using the Shapiro-Wilk test [29], we could not refute the hypothesis these variables were normally distributed.

**Table 2** – RBC parameters determined by DM in control and SCD groups.

|  | Control (N=8) | SCD (N = 9) | P-value |
|---|---|---|---|
| **Radius (µm)** | 3.93 [3.70 – 4.04] | 4.15 [3.77 – 4.68] | 0.0598 |
| **Maximum thickness (µm)** | 2.03 [1.65 – 2.23] | 2.03 [1.54 – 2.36] | 0.9740 |
| **Surface Area (µm$^2$)** | 126.4 [112.4 – 136.7] | 133.8 [98.5 – 159.5] | 0.3459 |
| **Volume (µm$^3$)** | 83.1 [65.1 – 95.6] | 83.2 [54.7 – 100.5] | 0.9884 |
| **Sphericity index** | 0.73 [0.70 – 0.76] | 0.69 [0.65 – 0.71] | **0.0003** |

Comparing control and SCD groups, there was no significant different in RBC radius, maximum thickness, surface area, and volume (**Table 2**). RBC sphericity index was significantly lower in SCD when compared with control group (0.69 [0.65 – 0.71] versus 0.73 [0.70 – 0.76]; P = 0.0003). Such difference contributes to a lower RBC deformability in SCD, since deformability is directly related with sphericity index [37]. These findings are corroborated by studies that associated SCD with lower RBC deformability, resulting in significant alterations of the microcirculatory blood flow [38,39].

Lower RBC sphericity index was also associated with more severe retinopathy. The area under the ROC curve to classify severe PSCR among individuals with SCD was 0.85 [0.53 – 1.00]. RBC sphericity index was also significantly lower in severe PSCR than in mild PSCR (0.67 [0.65 – 0.69] versus 0.70 [0.68 – 0.71]; P = 0.0420). The estimated threshold of mean sphericity index to predict severe PSCR was 0.69. The proportion of participants with sphericity index ≤ 0.69 was significantly greater in severe PSCR than in mild PSCR (5/5 = 100% versus 1/4 = 25%; OR: undefined; 95%CI: > 1.03; P = 0.0476). Using this threshold for RBC sphericity index, we achieved 100% sensitivity [55% - 100%] and 75% specificity [24% - 99%] to detect severe PSCR among participants with SCD. Histograms of RBC sphericity index for all study subgroups are shown in **Figure 5**.

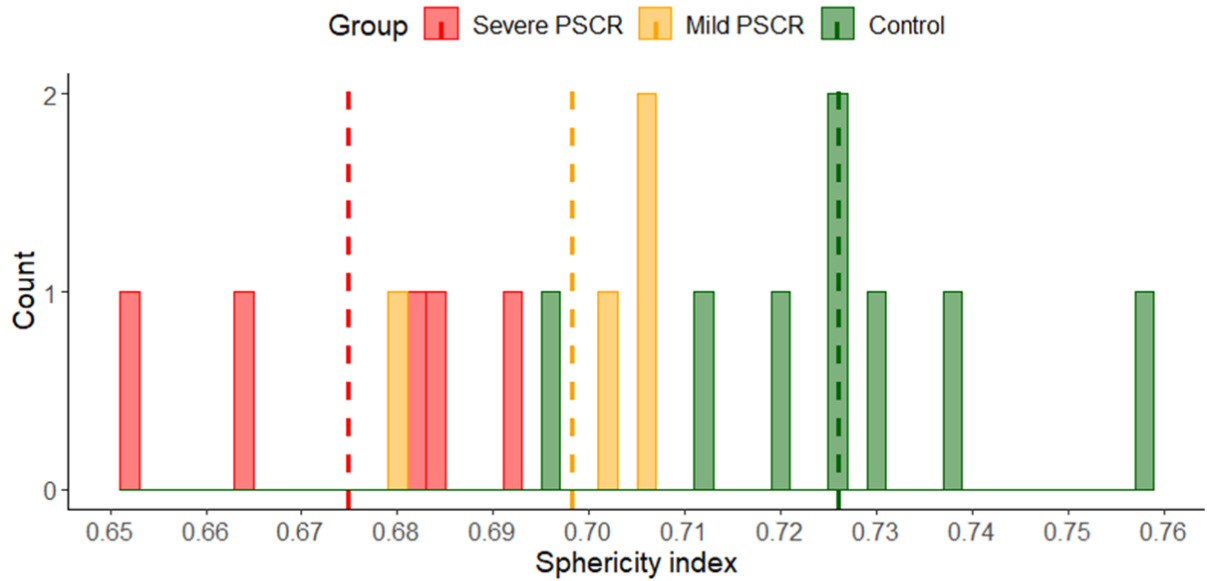

**Figure 5** – Histograms of average RBC sphericity index from individuals in control (green), mild (yellow) and severe (red) PSCR subgroups. Vertical dashed lines indicate mean values in each subgroup. Mild PSCR: proliferative sickle cell retinopathy (Goldberg < III). Severe PSCR: proliferative sickle cell retinopathy (Goldberg ≥ III).

A possible confounding factor in this study was hydroxycarbamide use in 3 participants with SCD (33%). As stated before, this drug increases the sphericity index of RBCs [36,37]. This effect would be greater especially in the severe PSCR group, in which 40% of participants were using hydroxycarbamide (versus 25% in the mild PSCR group). Nevertheless, a lower RBC sphericity index was observed in severe PSCR, suggesting an even stronger association between lower sphericity index and more severe retinopathy in SCD.

It has been shown that RBC deformability, associated with the sphericity index, correlates negatively with hemolysis levels in SCD [40,41]. For sickle cell retinopathy, however, the known risk factors, such as HbSC subtype, older age, male gender, lower levels of fetal hemoglobin (HbF), and higher hemoglobin concentration, could not be consistently used to predict more severe disease [3-7]. The current recommendation of annual ophthalmic screening in patients with SCD, for example, lacks strong supporting data and was defined by expert consensus [42]. In this study, we show that RBC sphericity index obtained by DM may represent a potential biomarker of PSCR severity.

DM is a simple bright field microscopy technique that requires a very small volume of blood (0.5 µL) and can also be applied to capillary blood samples, without the need of anticoagulants [21]. It can be useful for understanding and possibly for stratifying the risk of several SCD complications, as it determines morphological and mechanical characteristics of unlabeled RBCs (single-cell level) that are intrinsically related with cell deformability and blood rheology, such as RBC sphericity index and membrane fluctuation. Membrane fluctuation may also be related with PSCR severity, since it is another factor that influences RBC deformability. However, we did not include this parameter in this study because analysis would be underpowered; we estimated that at least 25 individuals would be necessary in each group to allow comparisons of membrane fluctuation with 90% power and 5% significance levels [28]. Nevertheless, this parameter is also automatically calculated by our algorithm (**Supplementary code**).

The high complexity and heterogeneity in methodology of previous studies on RBC deformability (e.g optical tweezers [43,44] and ektacytometry [45,46]) limit the comparison between different works, as well as clinical validation of reported findings. Additionally, several imaging methods are subjective [47]. We contribute to overcome these limitations making the developed algorithm available, facilitating study of RBC characteristics at single-cell level, in an objective way, in more individuals and in diverse diseases.

This study has limitations. Despite the adequate power to study associations between sphericity index and PSCR, the sample was too small to study associations of membrane fluctuation and PSCR. Only HbSC individuals were included in the SCD group because this is the subtype with highest risk of PSCR. However, including HbSS individuals would have been interesting to understand differences between these subtypes, which present different clinical manifestations and disease severity. Finally, we did not consider some factors that may be associated with PSCR severity, such as HbF levels, genetic polymorphisms, and RBC adhesive properties [1,48].

To conclude, we developed and made public an algorithm to automatically analyze multiple DM images taking about seven seconds per RBC analyzed. Lower RBC sphericity index, determined by DM, was associated not only with the presence of SCD, but also with more severe retinopathy, possibly serving as a biomarker of disease severity.

**Author contributions**

CB-d-R recruited participants to the study, conducted the experiments, wrote the computational code, did the statistical analysis, wrote the main manuscript text, and prepared the figures. UA, ONM and LSG participated in the study design, in the experimental setup and in writing the code. DVV-S participated in the study design and in recruiting subjects to the study. All authors reviewed the manuscript.

**Supplementary Equations**

This derivation of defocusing microscopy (DM) equations was adapted from the works of Mesquita et al. and Roma et al. [21,22]. The time mean intensity contrast $\langle C \rangle$ and time mean square intensity contrast fluctuation $\langle (\Delta C)^2 \rangle$ for each pixel captured at a specific focus distance $z_f$ can be calculated as:

$$\langle C(\vec{\rho}, z_f) \rangle = \frac{\langle I(\vec{\rho}, z_f) \rangle - I_0}{I_0}; \quad \langle (\Delta C)^2 \rangle = \frac{\langle I^2(\vec{\rho}, z_f) \rangle - \langle I(\vec{\rho}, z_f) \rangle^2}{I_0^2}; \quad (1)$$

where $\vec{\rho}$ is the two-dimensional position vector of a point at cell plane, $\langle I(\vec{\rho}, z_f) \rangle$ is the mean square intensity, $I_0$ is the time mean for background intensity and $\langle I^2(\vec{\rho}, z_f) \rangle$ is the time mean square intensity. The DM contrast for an object with two diffracting surfaces, at small defocusing distances ($z_f = \pm 2\mu m$ for a RBC), is:

$$C(\vec{\rho}, z_f) = \frac{\Delta n}{n_0} \{[z_f - h_1(\vec{\rho})]\nabla^2 h_1(\vec{\rho}) + [z_f + |h_2(\vec{\rho})|]\nabla^2 |h_2(\vec{\rho})|\}; \quad (2)$$

where $\Delta n$ is the refractive index difference between the RBC and the surrounding solution, $n_0$ is the immersion oil refractive index, and $h_1(\vec{\rho})$ e $h_2(\vec{\rho})$ are the height functions for upper and lower RBC surface membranes, respectively. Subtracting, pixel by pixel, contrast images (average of 300 frames) taken at $z_{f0} = 0\mu m$ and $z_{f2} = +2\mu m$:

$$\langle C_{z_{f2}} \rangle - \langle C_{z_{f0}} \rangle = \frac{\Delta n}{n_0} (z_{f2} - z_{f0}) \nabla^2 H(\vec{\rho}); \quad (3)$$

where $H(\vec{\rho}) = h_1(\vec{\rho}) + |h_2(\vec{\rho})|$ is the RBC thickness map. If we sequentially apply a Fourier transform to equation (3), divide it by $-q^2$ (where $q$ is the number of pixels in each image dimension), and apply an inverse Fourier transform, we can recover the RBC thickness map $H$:

$$H = \frac{n_0}{\Delta n(z_{f2} - z_{f0})} \mathcal{F}^{-1} \left( \frac{\mathcal{F}\{\langle C_{z_{f2}} \rangle - \langle C_{z_{f0}} \rangle\}}{-q^2} \right). \quad (4)$$

From the thickness map $H$, we can get the RBC volume from the pixel area $A_{pixel}$ and the RBC thickness in each pixel $H_{pixel}$:

$$VOLUME = \sum_{pixel} A_{pixel} \, H_{pixel}. \quad (5)$$

The RBC surface area (AREA), assuming RBC symmetry in relation to the horizontal plane, and the sphericity index (SPHERICITY) are:

$$AREA = 2 A_{pixel} \sum_{pixel} \sqrt{1 + \left(\frac{\nabla_x H_{pixel}}{2}\right)^2 + \left(\frac{\nabla_y H_{pixel}}{2}\right)^2}; \qquad (6)$$

$$SPHERICITY = 4,838 \frac{VOLUME^{2/3}}{AREA}. \qquad (7)$$

Membrane mean square fluctuation amplitude $\langle u^2 \rangle$ is proportional to the mean square intensity contrast fluctuation $\langle (\Delta C)^2 \rangle$, for $z_f > +3\mu m$, where $k_0$ is the illumination light wavenumber:

$$\langle \Delta C^2_{z_f \to \infty} \rangle = (\Delta n k_0)^2 \langle u^2 \rangle. \qquad (8)$$

Finally, we can estimate $\Delta n$ from the mean cell hemoglobin concentration (MCHC), obtained from a complete blood count:

$$\Delta n = 0,002 \, MCHC. \qquad (9)$$